\documentclass[11pt]{article}

\usepackage[margin=1in]{geometry}
\usepackage{tabularx}
\usepackage{booktabs}
\usepackage{multirow}
\usepackage[table]{xcolor}
\usepackage{tikz}
\usepackage{graphicx}
\usepackage{hyperref}

\definecolor{model}{HTML}{a00e00}
\definecolor{metric}{HTML}{d04e00}
\definecolor{task}{HTML}{0086a8}
\definecolor{dataset}{HTML}{f6c200}

\title{\textbf{Research Opportunities and Challenges of the EU’s Digital Services Act}}

\author{
Francesco Pierri$^{1,\dagger,*}$,\\
Theo Araujo$^{2}$,
Sanne Kruikemeier$^{3}$,
Philipp Lorenz{-}Spreen$^{4,5}$,\\
Mariek M. P. Vanden Abeele$^{6}$,
Laura Vandenbosch$^{7}$,\\
Joana Gon\c{c}alves-Sa$^{8,9,\dagger,*}$, 
Przemyslaw A. Grabowicz$^{10,11,\dagger,*}$
}

\date{
\small
$^{1}$Dipartimento di Elettronica, Informazione e Bioingegneria, Politecnico di Milano, Italy\\
$^{2}$Amsterdam School of Communication Research (ASCoR), University of Amsterdam, Netherlands\\
$^{3}$Wageningen University \& Research, Netherlands\\
$^{4}$Center Synergy of Systems, Dresden University of Technology, Germany\\
$^{5}$Max Planck Institute for Human Development, Germany\\
$^{6}$IMEC-MICT, Ghent University, Belgium\\
$^{7}$Media Psychology Lab, University of Leuven, Belgium\\
$^{8}$SPAC – LIP, Laboratory of Particle Physics, Portugal\\
$^{9}$NOVA LINCS, Portugal; External Faculty at Complexity Science Hub, Austria\\
$^{10}$University College Dublin, Ireland\\
$^{11}$University of Massachusetts Amherst, USA\\[4pt]
$^{\dagger}$Equally contributing authors\\
$^{*}$Corresponding authors: \texttt{francesco.pierri@polimi.it}, \texttt{joanagsa@lip.pt}, \texttt{przemek.grabowicz@ucd.ie}\\[4pt]
\smallskip
\textcolor{red}{This is the pre-print version of an article that will appear in \textbf{Communications of the ACM}. Please cite the published version when referencing this work.}
}

\begin{document}

\maketitle
\thispagestyle{empty}

\section*{Abstract}
The Digital Services Act (DSA) introduced by the European Union in 2022 offers a landmark framework for platform transparency, with Article 40 enabling vetted researchers to access data from major online platforms. 
Yet significant legal, technical, and organizational barriers still hinder effective research on systemic online risks. 
This piece outlines the key challenges emerging from the Article 40 process and proposes practical measures to ensure that the DSA fulfills its transparency and accountability goals.

\section*{Introduction}
Article 40 of the Digital Services Act (DSA), enacted by the European Union (EU) in 2022, aims to improve platform transparency by enabling independent researchers to study electoral interferences, algorithmic biases, privacy issues, and public health implications. 
To this end, Article 40 allows vetted researchers to access platform data. 
Significant efforts have been made in the past years to implement this process.
However, as of 2025, when the Delegated Act on Article 40 was adopted, it has become evident that significant barriers remain in practice. This piece was written collectively by a group of researchers with direct experience in EU policy discussions on the Delegated Act, namely, a 18-month European Research Council (ERC)-supported pilot program that brought together researchers, the DG-CONNECT, and national Digital Services Coordinators (DSCs),\footnote{Recommendations were presented during a public workshop on October 2025: \url{https://erc.europa.eu/news-events/events/erc-workshop-data-access-under-digital-services-act-dsa-article}} and a roundtable discussion with major platforms and the European Commission.
We provide an informed perspective on Article~40, discussing both its significance for scientific research and the obstacles -- such as conflicts of interest -- that must be addressed to ensure its success. We propose practical measures, including the development of standardized data-sharing frameworks, clear vetting procedures, and stronger collaboration between researchers and platforms to mitigate risks and increase transparency.

\section*{Opportunities}
The widespread use of very large online platforms (VLOPs) and search engines (VLOSEs) has serious and far-reaching consequences for society.\footnote{\url{https://digital-strategy.ec.europa.eu/en/policies/digital-services-act-package}} Recognizing the importance of regulating such platforms, the DSA identifies four categories of systemic risks to society from online platforms -- risks to democratic processes, fundamental rights, public health, and the dissemination of illegal content~\cite{Lorenz} -- with obligations for platforms to minimize them. Article 40 of the Act marks a milestone for enhancing transparency and accountability: it provides vetted researchers with a legal framework and a technical pipeline to request access to public and non-public data from VLOPs and VLOSEs. Both data types are essential to identify, study, and offer potential mitigation strategies for the aforementioned systemic risks.

Data access can help researchers detect and monitor serious instances of electoral interference in democratic processes through online disinformation and opaque social media algorithms.\footnote{An example of a study investigating potential biases in the feed algorithm of X before the 2024 German federal election: \url{https://doi.org/10.5281/zenodo.14894900}.} Regarding fundamental rights, it may help identify algorithmic designs that are biased toward or against particular societal groups~\cite{Ali}. Similarly, it could also help to evaluate platforms’ compliance with critical regulations, such as the GDPR and the DSA itself, addressing issues such as privacy violations~\cite{Isaak}. Finally, it provides researchers with tools to study public health risks, including access to information during health emergencies, the effects of social media on the mental and physical well-being of their users (particularly among young and vulnerable ones)~\cite{Valkenburg}, and to devise strategies to combat the dissemination of illegal content and disinformation. However, many challenges remain.

\section*{Misaligned Incentives}
First and foremost, the goals of researchers, EU citizens, national governments, and online platforms do not always align and sometimes even clash, resulting in reduced transparency. Online platforms prioritize protecting the interests of their owners or stakeholders, and it is increasingly clear that researchers should expect resistance from them--from delays and bureaucratic or technical hurdles to legal obstruction. In meetings with several platforms, their representatives have expressed different levels of reluctance to the direct sharing of data, instead favoring a range of alternative approaches. One suggestion was to follow the ``independence by permission'' model in which the researchers state the hypotheses and the platforms test them~\cite{Wagner}. However, industry research often produces outcomes aligning with their own interests~\cite{LopezMoreno}. For instance, a recent independent study that analysed research produced by a collaboration between Meta and two dozen prominent researchers~\cite{Bagchi}, found that the research portrayed Meta's platforms as more effective at curbing misinformation than they are in normal circumstances by not revealing temporary algorithmic changes deployed at the time of the study due to the 2020 US presidential election.\footnote{A comment about this incident from the chief editors of Science and Meta representatives: \url{https://www.science.org/doi/10.1126/science.adt2983}} The official independent observer of this collaboration concluded that regulation allowing independent research, such as the DSA, is indispensable~\cite{Wagner}. During our meetings, some other platforms alluded to potential legal action against researchers and NGOs, with already set precedents.\footnote{\url{https://www.reuters.com/technology/musks-x-corp-loses-lawsuit-against-hate-speech-watchdog-2024-03-25/}}\textsuperscript{,}\footnote{\url{https://www.politico.eu/article/x-challenges-german-court-decision-that-would-force-it-to-share-data-with-researchers/}} Most online platforms are black boxes by design, and in recent years they have been increasingly restricting data access amid a global AI race, to prevent competing AI providers from training on their data without permission. The alignment of online platforms with the current US government,\footnote{\url{https://www.politico.eu/article/us-congress-eu-digital-services-act-foreign-censorship/}}
involving the gradual cancellation of fact-checking programs\footnote{\url{https://www.science.org/doi/full/10.1126/science.adv4632}} and diversity, equity, and inclusion (DEI) initiatives, coupled with the government's increasing cuts and halts in funding for these topics,\footnote{\url{https://www.nature.com/articles/d41586-025-00562-w}} reflect a broader trend of limiting independent scrutiny and oversight. The DSA reflects Europe’s markedly different approach, setting a precedent for responsible governance in the digital space and underscoring the importance of independent research for online risk-prevention~\cite{Liesenfeld}.

\section*{Power and Resource Asymmetries}
Another major challenge to DSA's implementation is the gap between academic and corporate research in access to various types of resources. First, rapidly growing AI industry research has overshadowed public and fundamental research financially and infrastructurally~\cite{Ahmed,Shekhtman}. Second, there is an asymmetry in information access: platforms know what they have, but researchers do not. Third, platforms have sizable and well-resourced legal teams, whereas university legal teams are small and fragmented. Fourth, platforms are increasingly used, even by their owners, to promote self-serving narratives that challenge open science and attack worldwide transparency regulation, including the DSA, by framing it as censorship\footnote{\url{https://www.reuters.com/technology/we-do-not-censor-social-media-eu-says-response-meta-2025-01-08/}} or by threatening with tariffs to be used as leverage in trade deals.\footnote{\url{https://www.ft.com/content/8f38514b-3265-496e-91e2-f8a44daa0ab6}} These social media campaigns already reached tens of millions of people and had a large impact.\footnote{\url{https://www.theguardian.com/world/2025/jan/08/eu-commission-urged-to-act-elon-musk-interference-elections}} In contrast, academic researchers have no means to access similar computing and data infrastructure, engage in long legal battles, or reach such large public audiences, to counterbalance these one-sided narratives. For example, political pressure and legal costs have led to the dismantling of entire research labs.\footnote{\url{https://www.platformer.news/stanford-internet-observatory-shutdown-stamos-diresta-sio}} These issues can further diminish trust in science and hamper its ability to inform society.

\section*{Implementation Bottlenecks}
There are also several practical challenges to balancing privacy, platform security, and the principles of transparency, reproducibility, and open science. Article 40, together with its Delegated Act adopted in July 2025, provides a mechanism for addressing these challenges. The European Commission, particularly through DG-Connect, has created a procedure to facilitate data access, enabling researchers to make data requests to VLOPs and VLOSEs while protecting their security and trade secrets, as well as user privacy. These requests can now be completed through a dedicated portal\footnote{\url{https://algorithmic-transparency.ec.europa.eu/news/faqs-dsa-data-access-researchers-2025-07-03\_en}} and the first decisions are expected in late February 2026. However, based on our direct involvement with the current processes around this pipeline, we recognize that significant efforts are still needed. It remains vital to inform and support the community in navigating the opportunities and procedures established by the DSA. Moreover, interactions between researchers and industry will be overseen by national Digital Service Coordinators (DSCs) and local university Data Protection Officers (DPOs), and researchers will undergo independent vetting and disclose funding sources, potential conflicts of interest, and plans for data handling and processing. This process is likely to be slow, particularly for the first requests as there will likely be a mismatch in language and expectations between researchers, the DSCs, DPOs, and the platforms. Researchers typically focus on empirical inquiries, scientific method, transparency, and academic independence, while regulators and officers are more often preoccupied with legal liability and risk-aversion. This disconnect can lead to misunderstandings and the imposition of unnecessary restrictions, even when researchers are operating in full compliance with ethical and legal standards. Equally important will be safeguards to protect vetted researchers from undue influence, promoting effective and seamless exchange of knowledge. Compounding these structural issues is a circular problem of research resources. On the one hand, access to data is often required as a prerequisite to submitting competitive research proposals and securing grants; on the other, without dedicated funding, many researchers -- especially independent, early-career, or under-resourced ones -- lack the time and legal support to endure the complex, iterative process of applying for data access, creating a vicious cycle: those without data cannot secure funding, and without funding, they cannot obtain data.  

\section*{Recommendations and Conclusion}
To achieve a successful implementation, we make three key recommendations that directly stem from the described challenges, particularly in reducing the asymmetries in resources:
\begin{enumerate}
\item \textbf{Streamline access procedures}: There is an urgent need for the development of processes that will both reduce the costs for all stakeholders, and the processing time of data access requests. Efficient processes should be grounded in systematic categorisation of datasets and their sensitivity levels, and adoption of standards for data sharing agreements, documentation, modalities, and safeguards, which could be provided through the DSA data access portal. In its initial years, this may be developed gradually, leveraging existing datasets and extending them to cover their gaps and to unify them across platforms, as well as potentially reviving discontinued ones. An effective implementation of the DSA will likely require coordinated training of national DSCs and local university DPOs to prepare them for processing requests for various kinds of platform datasets, often massive in size and containing aggregated data of users.
\item \textbf{Support and foster independent research}: The outcomes of DSA-enabled research will be instrumental to inform official audits and policymaking but, so far, there is no planned support nor dedicated mechanisms to address the mentioned circular problem of resources. Such instruments should be envisioned. In parallel, it is essential that academics, even those typically more hesitant about regulation, understand the significance and urgency of the DSA and contribute to its effective implementation. Researchers should coordinate their efforts -- including sharing best practices, lessons learned, and successful approaches -- and promote bottom-up initiatives, from tutorials to outreach actions, to build collective knowledge and capacity, and strengthen public support for the Act. 
\item \textbf{Close regulatory blind spots}: This is a rapidly evolving field, with technological advancements consistently outpacing regulation, making it essential that frameworks adapt swiftly and proactively. We argue that large language models (LLMs), such as ChatGPT, meet all the criteria of the DSA: they are intermediary services deployed at scale, used by millions of EU users, and capable of inducing systemic risks, e.g., by influencing electoral processes in the EU. Increasingly, LLMs are integrated into search platforms (e.g., Bing with ChatGPT) and social media platforms (e.g., Grok on X), shaping how users access and interpret information. This technology may amplify the societal impact of conventional VLOPs and VLOSEs, so this regulatory blind spot should be urgently addressed.
\end{enumerate}
Finally, we urge the European Commission to ensure rigorous oversight of the DSA’s implementation, preventing industry resistance from weakening its impact, and call for the research community to coordinate around data access procedures. As the role of online platforms grows in importance, the successful implementation of the DSA is even more urgent now than at the time when it was developed. If the implementation of Article 40 is obstructed, researchers’ ability to scrutinize digital media would be severely limited, thereby undermining public trust and transparency. Europe's proactive stance should serve as a model for other regions and pave the way for a more transparent and safe digital ecosystem. 

\begin{figure}[!t]
    \centering
    \includegraphics[width=0.75\linewidth]{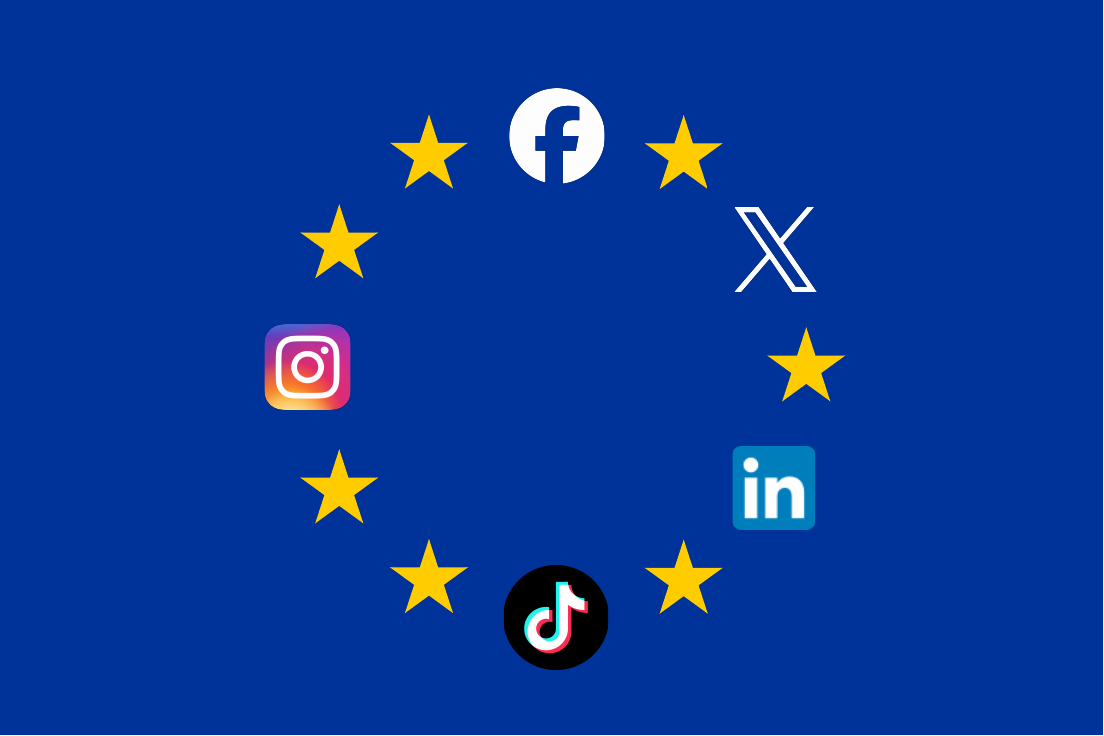}
    \label{fig:placeholder}
\end{figure}


\end{document}